\documentclass[fleqn,10pt]{wlscirep}
\usepackage[utf8]{inputenc}
\usepackage[T1]{fontenc}
\usepackage{array,multirow}
\usepackage{graphicx}
\usepackage{subcaption}

\title{Automated Estimation of Total Lung Volume using Chest Radiographs and Deep Learning}

\author[1,*]{Ecem Sogancioglu}
\author[1]{Keelin Murphy}
\author[1]{Ernst Th. Scholten}
\author[1]{Luuk H. Boulogne}
\author[1]{Mathias Prokop}
\author[1]{Bram van Ginneken}

\affil[1]{Radboud university medical center, Institute for Health Sciences, Department of Medical Imaging, Nijmegen, The Netherlands}
\affil[*]{ecem.sogancioglu@radboudumc.nl}

\begin{abstract}
Total lung volume is an important quantitative biomarker and is used for the assessment of restrictive lung diseases. In this study, we investigate the performance of several deep-learning approaches for automated measurement of total lung volume from chest radiographs. 7621 posteroanterior and lateral view chest radiographs (CXR) were collected from patients with chest CT available. Similarly, 928 CXR studies were chosen from patients with pulmonary function test (PFT) results. The reference total lung volume was calculated from lung segmentation on CT or PFT data, respectively. This dataset was used to train deep-learning architectures to predict total lung volume from chest radiographs. The experiments were constructed in a step-wise fashion with increasing complexity to demonstrate the effect of training with CT-derived labels only and the sources of error. The optimal models were tested on 291 CXR studies with reference lung volume obtained from PFT. The optimal deep-learning regression model showed an MAE of 408 ml and an MAPE of 8.1\% and Pearson's r = 0.92 using both frontal and lateral chest radiographs as input. CT-derived labels were useful for pre-training but optimal performance was obtained by fine-tuning the network with PFT-derived labels. We demonstrate, for the first time, that state-of-the-art deep learning solutions can accurately measure total lung volume from plain chest radiographs. The proposed model can be used to obtain total lung volume from routinely acquired chest radiographs at no additional cost and could be a useful tool to identify trends over time in patients referred regularly for chest x-rays.
\end{abstract}
\begin{document}

\flushbottom
\maketitle
\thispagestyle{empty}

\section*{Introduction}
Chest radiography (CXR) remains the most commonly performed imaging technique and one of the most often repeated exams because of its low cost, rapid acquisition and low radiation exposure \cite{Raoo12}. It was estimated that 129 million chest radiographs were performed in 2006 in the United States alone \cite{Mett09}. Chest radiographs play an important role in screening, monitoring, diagnosis, and management of thoracic diseases.

Wide availability of CXR has motivated researchers to build artificial intelligence (AI) systems that can automatically detect a variety of abnormalities \cite{Anna19,Rajp18,Murp20} and extract quantitative clinical measurements from them \cite{Sogan20,Li19}. AI systems have potential use for routine quantification of numerous biomarkers related to lung diseases, cardiac health, or osteoporosis. Applying such systems, whenever a chest radiograph is acquired, would be a step towards routine quantitative radiology reporting. 

This work focuses on an important quantitative biomarker, total lung volume, and investigates whether it can be measured automatically from plain chest radiographs using state-of-the-art deep learning approaches. Total lung volume (TLV) is used for assessing severity, progression and response to treatment in restrictive lung diseases \cite{Quan93,Flesc12}. Specific temporal changes in TLV can be identified in patients with obstructive and restrictive lung diseases, such as emphysema, pulmonary fibrosis or asthma. Further, TLV has been shown to correlate with mortality and health status \cite{Pedo12}.

Currently, the gold standard for measurement of TLV is the pulmonary function test (PFT), using special techniques such as body plethysmography, helium, or nitrogen dilution techniques \cite{Flesc12}. Several studies \cite{Tantu16,Coxs08,Iwan09} demonstrated that TLV measured from CT strongly correlates to TLV obtained from PFTs. Alternatively, several studies investigated TLV estimation from CXR using predictive equations. In fact, this has been a research interest for a century, with the first paper appearing in 1918 (\cite{Lund18}) demonstrating the correlation of external measurements from CXR to the pulmonary function test (gas dilution technique). All such previous literature, investigating predictive equations, was either based on the use of planimetric techniques \cite{ries89,Harr71,Cobb54,Schl95}, or made assumption of a given a geometry \cite{Barn60,Pier79,Loyd66}, or required several manual linear measurements to estimate TLV from CXR. However, all these studies required manual measurements to estimate TLV and used small sample sizes, making it unclear whether the techniques could be generalized to other populations.

In this study, we investigate, to the best of our knowledge, for the first time, whether chest radiography can be used to automatically predict TLV in a fully automated fashion using large datasets and deep learning. We examine the role of TLV labels derived from thoracic CT imaging in training deep learning systems. In order to account for variations in inspiration and dataset complexity, experiments with simulated and real chest radiographs in three different datasets were designed in a step-wise fashion. For each experiment, we optimized various state-of-the-art deep learning regression approaches to predict TLV using only posterioranterior (PA) view, lateral view or both views. The purpose of our study was to determine the accuracy of fully automatic measurement of TLV from CXR using deep learning based models. 

\section{Materials and Methods}
\subsection{Data and Preprocessing}
The data used in this study was obtained from two sources; the COPDGene study~\cite{Rega10} and Radboud University Medical Center (RUMC). To facilitate our stepwise experimentation, demonstrating sources of error, we experimented with simulated CXR images (digitally reconstructed radiographs), which are obtained from average intensity projections (AIP) on thoracic CT, as well as with true CXR images.  Reference total lung volume labels were obtained by two means; through segmentation of the lungs in CT and from pulmonary function tests (PFT).  The datasets constructed are described in detail in the sections below and in Figure~\ref{fig:diagram_data}. 

\subsubsection{COPDGene-sim}\label{sec:data:copdgene}

Inspiration chest CT studies (1000) from unique patients were randomly selected from the COPDGene study,  \cite{Rega10} which is publicly available on request for research purposes. The images in this study are acquired from patients with Chronic Obstructive Pulmonary Disorder (COPD), varying from mild to very severe. From the 1000 randomly selected CT studies, 800 (600 for training and 200 for validation) were used for training and validation, and 200 were retained as a held-out test set as illustrated at the top of Figure~\ref{fig:diagram_data}.

Lung segmentations were obtained by an automated algorithm and manually corrected by trained analysts with radiologist supervision \cite{Xie20}. Reference TLV was calculated for each CT scan by multiplying CT image spacing by the number of voxels segmented.

Simulated CXRs were generated from CT by creating AIP \cite{Camp18} from coronal and sagittal planes, resulting in frontal and lateral view simulated CXR. This dataset, which we refer to as COPDGene-sim, was used to demonstrate model performance in an ideal scenario where there is no inspiration difference between the label source (CT) and the (simulated) CXR image, CT segmentations are manually corrected, and the variety of pathologies is limited.

\subsubsection{RUMC Datasets}
This data was obtained from routine clinical care in Radboud University Medical Center, Nijmegen, the Netherlands (RUMC). Medical ethics committee approval was obtained prior to this study, and dataset was collected and anonymized according to local guidelines. 

\begin{figure}[!tbp]
    \centering
  \includegraphics[width=0.95\textwidth]{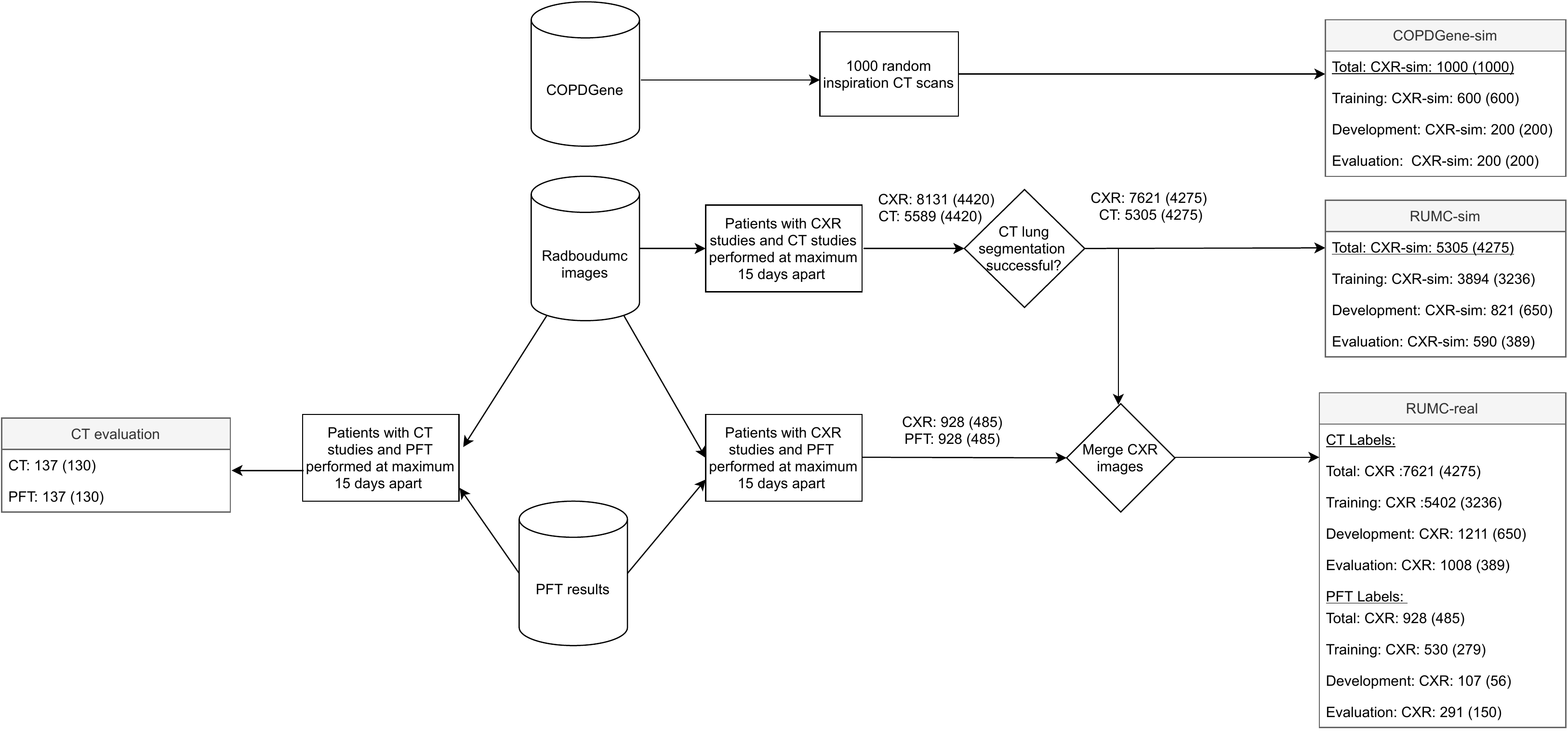}
  \caption{Flowchart that shows the criteria to select the data to be used in the experiments. Numbers of images are shown with numbers of patients in brackets. Abbreviations: CXR = chest radiographs, CXR-sim = simulated chest radiographs from CT, PFT = pulmonary function test.}
  \label{fig:diagram_data}
\end{figure}

We retrospectively collected CXR studies and chest CT acquired between 2003 and 2019 resulting in 321k CXR studies and 120k CT studies. Patients with both CT and CXR (with PA  and lateral view), performed a maximum of 15 days apart, were selected (4420 patients). The reference standard TLV measurements were obtained by a CT lung segmentation algorithm \cite{Xie20} and segmentation failure cases were visually identified and excluded (284 CT). This resulted in 7621 CXR studies and 5305 CT studies from 4275 patients (Figure \ref{fig:diagram_data}). Multiple CXR studies from a single patient could be matched to a single CT reference standard.

A group of patients being assessed for lobectomy was used to provide subjects with both PFT and CXR data acquired within 15 days of each other. This resulted in 928 CXR studies from 485 patients. Reference TLV was determined using the helium delusion technique \cite{Wang05}.

\begin{figure*}[h!]
\centering
\begin{subfigure}{0.26\textwidth}
\centering
    \includegraphics[width=\textwidth]{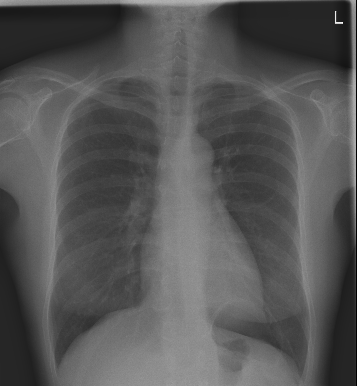}
    \caption{Real CXR}
    \label{fig:scattermodel}
\end{subfigure}
\begin{subfigure}{0.22\textwidth}
\centering
    \includegraphics[width=\textwidth]{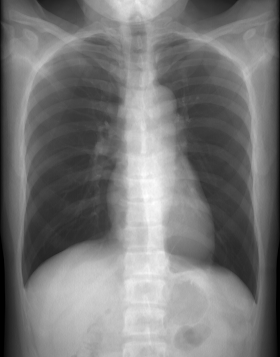}
    \caption{Fake CXR}
    \label{fig:scattersecond}
\end{subfigure}
\begin{subfigure}{0.49\textwidth}
\centering
    \includegraphics[width=\textwidth]{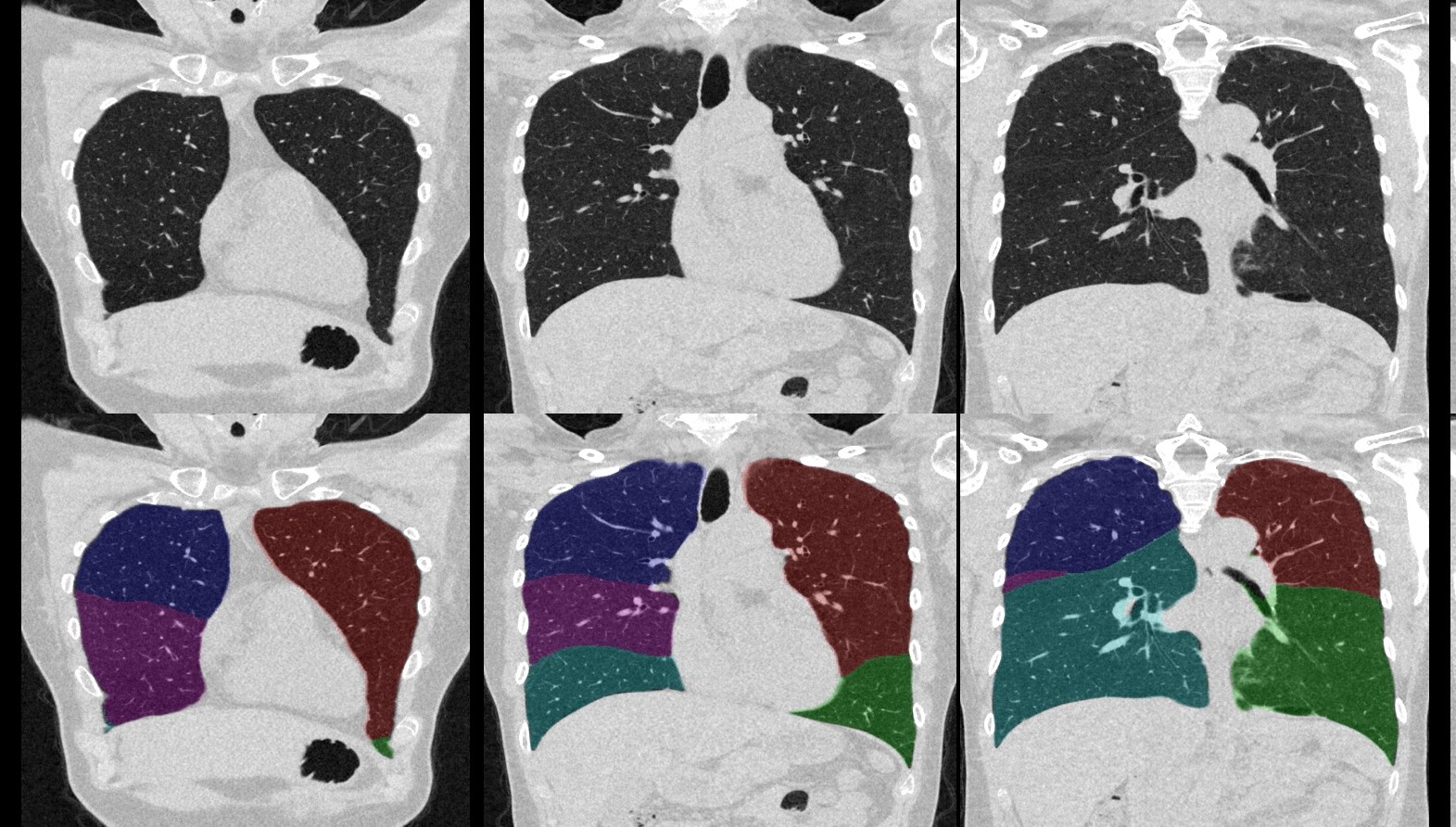}
    \caption{CT scan}
    \label{fig:histmodel}
\end{subfigure}\hfill
\caption{Real CXR (a), Simulated CXR (b) and coronal CT slices (c) from a patient in the RUMC-real dataset. Lobe segmentation results in CT are illustrated in the bottom row of (c). CT-derived TLV is calculated as the sum of the lobe volumes. CT-derived TLV for this subject was 3.8 liters, while PFT-derived TLV was 4.3 liters.}
\label{fig:exampleimages}
\end{figure*}

From this dataset, we created two sets for experimentation. The first is referred as RUMC-sim and used simulated CXR generated from CT as described in Section~\ref{sec:data:copdgene}. The second is RUMC-real, consisting of real CXR with CT-derived and PFT-derived labels for TLV. To investigate the relationship between CT-derived and PFT-derived labels, we created a dataset, CT-evaluation, where both CT and PFT were acquired within 15 days of each other. We made sure that there was no patient overlap between training and held-out evaluation sets for all the datasets. These datasets are detailed below and illustrated in Figures \ref{fig:diagram_data} and \ref{fig:exampleimages}.

\paragraph{RUMC-sim}{In this dataset, both frontal and lateral view chest radiographs were simulated from 5305 CT studies (4275 patients). Of these, 389 patients (590 CT studies) were randomly selected and used as a held-out evaluation set, whereas the remaining 3886 patients (3236 for training, 650 for validation) were used for training. This dataset, with CT-derived lung volume labels, was used to illustrate the model performance in a set of images with a large variety of abnormalities (compared to COPDGene-sim), e.g., pleural fluid, large masses, widespread interstitial abnormalities. The use of simulated CXR images removes any possibility of error related to inspiration effort, or patient position between the label source (CT) and the (simulated) CXR.}

\paragraph{RUMC-real}{This dataset consists of patients with real CXR studies (PA and lateral) and with lung volume reference standard measurements from two sources, namely CT and PFT. For CT-based data, the same patient partitioning was used as in RUM-sim, but using the CXR with the study time closest to that of the corresponding CT study rather than a simulated CXR. This resulted in 7621 CXR studies with CT-derived labels, whereas PFT-derived labels were used for 928 CXR studies as seen in Figure \ref{fig:diagram_data}. As a held-out evaluation set, 590 patients with 1008 CXRs with CT-derived labels, and 291 CXR from 150 patients with PFT-derived labels were randomly selected. We made sure there was no patient overlap between the PFT-based evaluation set and any training set (with CT-labels or PFT-labels).}

\paragraph{CT evaluation dataset}{We identified patients with PFT results available that were also in the RUMC-sim dataset, and selected patients with PFT results obtained a maximum of 15 days apart from their CT study. This resulted in 137 CT studies from 130 patients. CT lung volume was calculated by means of an automated CT lung segmentation algorithm \cite{Xie20}, and the results were visually inspected, identifying no obvious failed segmentations. This set was used to demonstrate the relationship between CT-derived and PFT-derived labels.}

All CT scans used in the COPDGene-sim and RUMC-sim datasets were first resampled to 1mm isotropic spacing before generating simulated CXRs by average intensity projection. Similarly, real CXRs were resampled to have 1mm $\times$  1mm spacing.
All real and simulated CXR images were padded with zeros to reach a fixed size of 512 x 512 pixels. Images underwent standard normalization to the range of -1 to 1.


\begin{table}[]
\begin{tabular}{|l|l|l|l|l|l|}
\hline
\multicolumn{2}{|l|}{} & \multicolumn{4}{c|}{Experimental Datasets} \\
\hline
\multicolumn{2}{|l|}{} & COPDGene-sim & RUMC-sim   & RUMC-real (CT-labels) & RUMC-real   (PFT-labels) \\ \hline
\begin{tabular}[c]{@{}l@{}}\end{tabular}                       & Label type                                                               & CT-derived   & CT-derived & CT-derived            & PFT-derived              \\ \hline
\multirow{4}{*}{\begin{tabular}[c]{@{}l@{}}Possible sources \\ \\ of label error\end{tabular}} & \begin{tabular}[c]{@{}l@{}}Patient position   \\ difference\end{tabular} &              &            & \textbf{Y}            & \textbf{Y}               \\ \cline{2-6} 
                                                                                               & \begin{tabular}[c]{@{}l@{}}Inspiration effort\\ difference\end{tabular}  &              &            & \textbf{Y}            & \textbf{Y}               \\ \cline{2-6} 
                                                                                               & \begin{tabular}[c]{@{}l@{}}CT segmentation\\ inaccuracy\end{tabular}     &              & \textbf{Y} & \textbf{Y}            &                          \\ \cline{2-6} 
                                                                                               & Diverse pathologies                                                      &              & \textbf{Y} & \textbf{Y}            & \textbf{Y}               \\ \hline
\end{tabular}
\caption{Datasets characteristics in step-wise experiments. RUMC-real (PFT-labels) was used to finetune the models which were pretrained on the RUMC-real (CT-derived) dataset. Y indicates that the condition holds true.}
\label{tab:dataset_analysis}
\end{table}

\subsection{Methods}\label{sec:methods}
We experiment with 5 different deep-learning architectures, 4 of which are widely used popular classification architectures (DenseNet121~\cite{Huan17}, ResNet34, ResNet50~\cite{He16}, VGGNet~\cite{Simo15}), and one, referred as 6-layer CNN, was designed to represent a shallow architecture. The 6-layer CNN consisted of 6 CNN layers, each followed by RELU, batch normalization and a pooling layer. The first CNN layer had 32 feature maps, and the number of feature maps was doubled in each layer.  The final CNN layer was followed by 3 fully connected layers which mapped the number of features to 512, 128 and 1, respectively. 

The dual CNN architecture, which receives both frontal and lateral radiographs as input, consists of two branches with a backbone architecture that is either VGG-Net, ResNet34 or 6-layer CNN, and concatenates the features from these branches before the first fully connected layer. Due to memory limitations, Densenet121, and ResNet50 architectures were not investigated for the dual CNN model.

\begin{figure}[!tbp]
    \centering
  \includegraphics[width=0.75\textwidth]{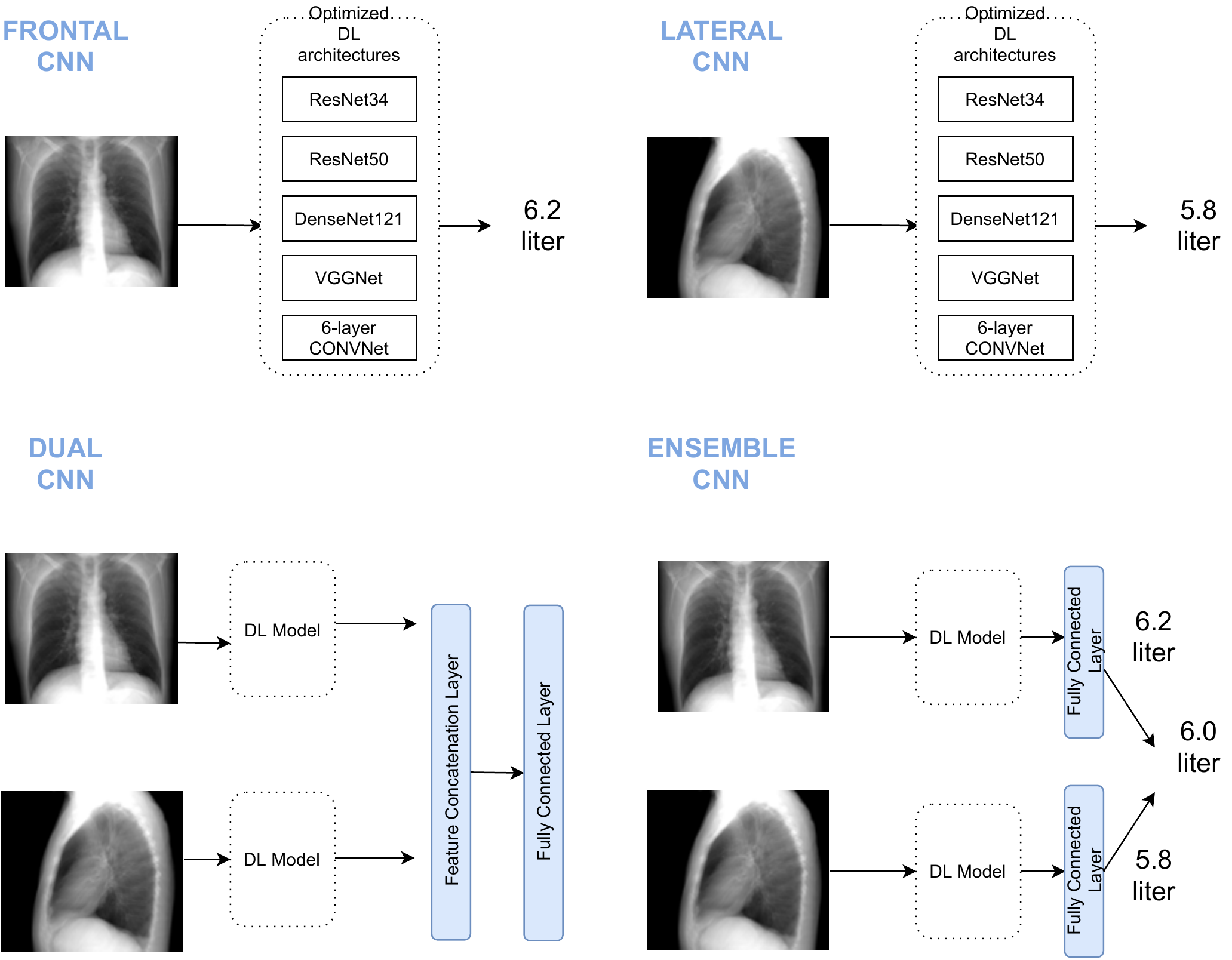}
  \caption{Illustration of architecture pipelines. Four different experimental designs were considered: frontal CNN, lateral CNN, dual CNN (combining frontal and lateral models by layer concatenation) and ensemble CNN (combining optimal frontal and lateral models by averaging their outputs).}
  \label{fig:model}
\end{figure}

These network architectures were trained with 3 possible inputs (PA CXR, lateral CXR or both, and methods of combining their outputs (see Figure \ref{fig:model}). Each network outputs a regression value representing TLV in liters. 

\begin{figure}[!tbp]
    \centering
  \includegraphics[width=0.8\textwidth]{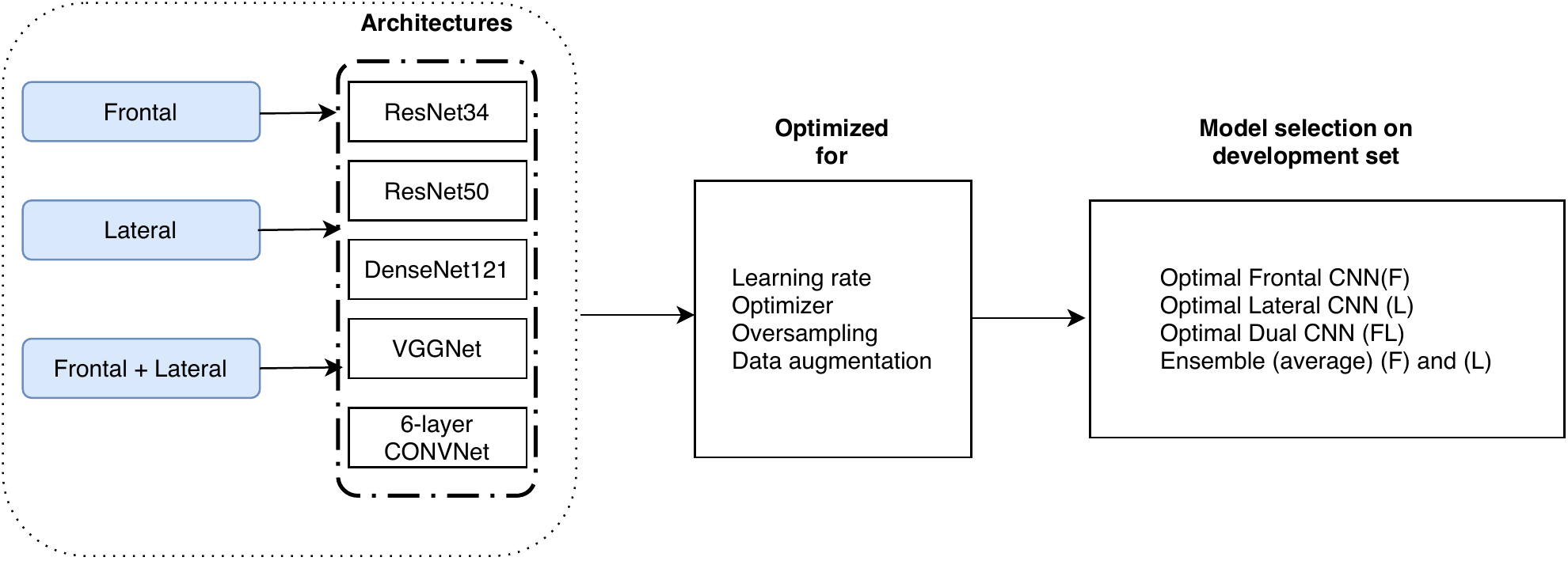}
  \caption{Illustration of our model selection process on validation set. Different network architectures were systematically optimized for three different inputs, namely frontal, lateral, and dual (frontal+lateral), separately. Each of them was optimized systematically for hyperparameters, and the model with the least mean absolute percentage error on the validation set was selected.}
  \label{fig:hyperparameteroptimization}
\end{figure}

For each model trained, a hyperparameter optimization was carried out to ensure the best possible result for that architecture/input combination on the validation set. A variety of aspects for training a convolutional neural network were considered as hyperparameters: they were learning rate, optimizer, oversampling technique, and data augmentation as seen in Figure \ref{fig:hyperparameteroptimization}. Random hyperparameter optimization was employed given a predefined range for hyperparameters for each model (frontal, lateral, and dual CNN) separately. 

Each model was trained by optimizing the mean squared error loss between the predicted TLV and the reference standard TLV. The model was trained for a maximum of 300 epochs, terminating if there was no improvement in the validation set performance for 50 successive epochs. We selected the epoch that yielded the least mean squared error in the validation set.

For each of our 3 datasets, the optimal combination of architecture and hyperparameters was identified for each of the 3 possible input types on the validation set. These models were then applied to the held-out evaluation set. In addition, the average of the 2 outputs from the networks using single (frontal or lateral) inputs is calculated and presented as Ensemble CNN output (Figure \ref{fig:model}).

Our TLV prediction experiments were constructed in a step-wise fashion, to identify potential sources of error as the task becomes increasingly difficult. This is illustrated in Table \ref{tab:dataset_analysis}. CT-derived volume labels are used in all experiments except the final one where the network is additionally fine-tuned with PFT-derived labels. We begin with the COPDGene-sim dataset, where errors related to patient position and inspiration effort as well as errors related to CT segmentation accuracy and diversity of underlying pathologies are eliminated. In RUMC-sim we introduce the potential for errors from minor CT segmentation inaccuracies, and from the diverse pathology within the dataset, which is likely to increase the variability in image appearance. Finally in RUMC-real, we first experiment with predicting CT-derived TLV from chest radiographs (RUMC-real (CT-labels)), and subsequently with PFT-derived TLV (RUMC-real (PFT labels)). In this last experiment, since there is only a small number of gold-standard PFT labels available (487 patients), the network trained with CT-labels is used as pre-trained model, and fine-tuned using CXR images with associated PFT-labels.

As an additional experiment, we investigate the relationship between PFT-derived TLV and CT-derived TLV, in a scenario where they are acquired at most 15 days apart from each other, using the CT-evaluation dataset. 

\subsection{Statistical Analysis}
Mean absolute error (MAE), mean absolute percentage error (MAPE) and Pearson correlation coefficient were computed to demonstrate the relationship between predicted and reference TLV values. The 95\% limits of agreement were estimated by means of a non-parametric method for Bland-Altman plot since the data distribution was not normal, as assessed with Shapiro-Wilk test \cite{Shap65} and quantile-quantile plot \cite{Wilk68}.

\section*{Results}
Model training for each model, namely frontal CNN, lateral CNN, and dual CNN, took between 8 to 14 hours on the RUMC-sim and RUMC-real datasets (depending on the network architecture), and 2 to 4 hours on the COPDGene-sim dataset using a variety of GPUs such as TitanX, GTX1080, GTX1080ti, GTXTitanx, and TitanV. The mean processing time per test image was 0.3 seconds.

Three trained models (frontal, lateral, dual) were selected for each dataset, based on optimization using the validation set, and applied to the held-out evaluation data. Additionally the outputs of the optimized frontal and lateral models were averaged and presented as “Ensemble” model. The selected architectures, and their performance on the held-out evaluation data are provided in Table \ref{tab:mae}. 

\begin{figure*}[h!]
\begin{subfigure}{0.49\textwidth}
\centering
    \includegraphics[width=\textwidth]{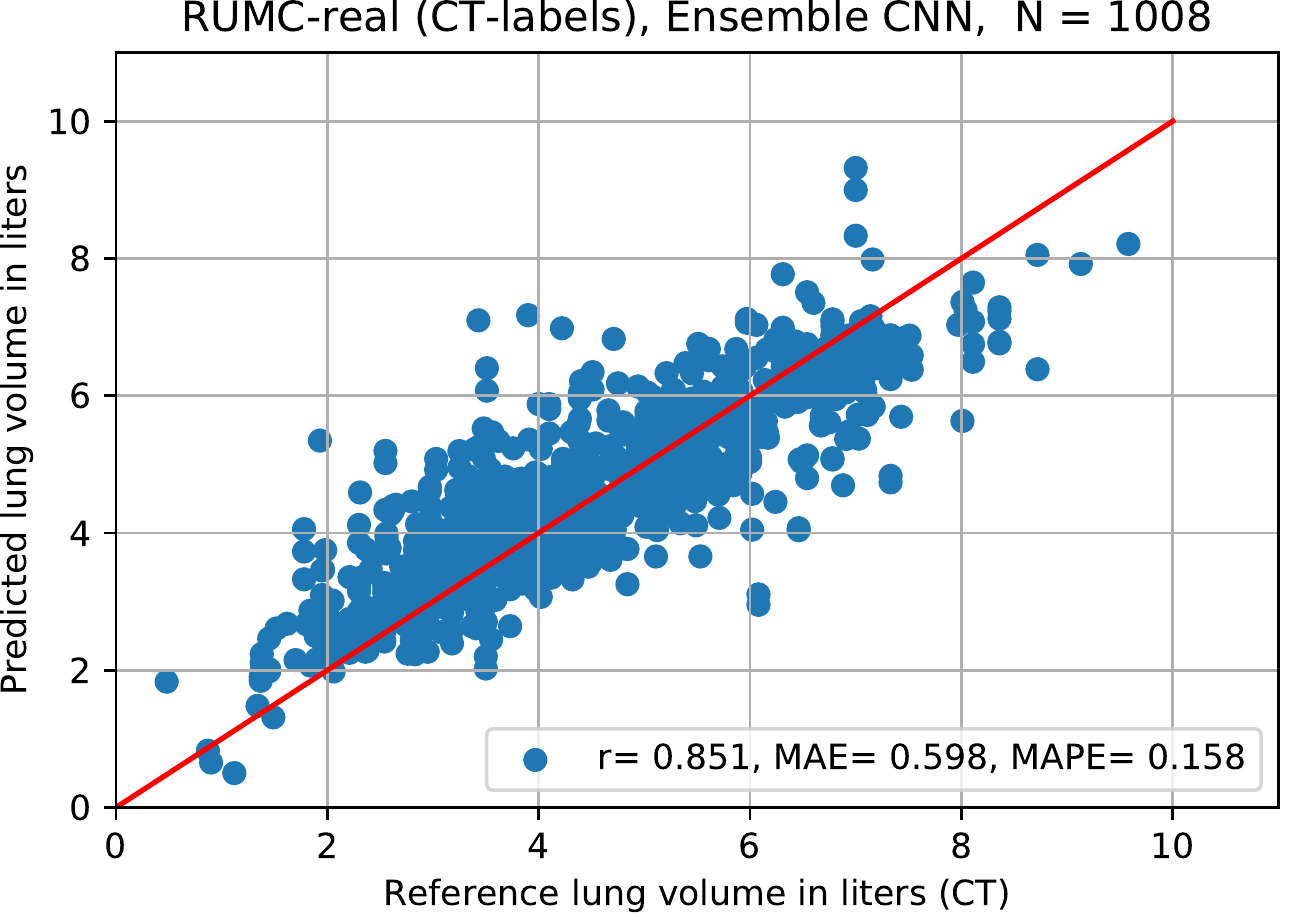}
    \caption{}

    \label{fig:histmodel}
\end{subfigure}\hfill
\begin{subfigure}{0.49\textwidth}
\centering
    \includegraphics[width=\textwidth]{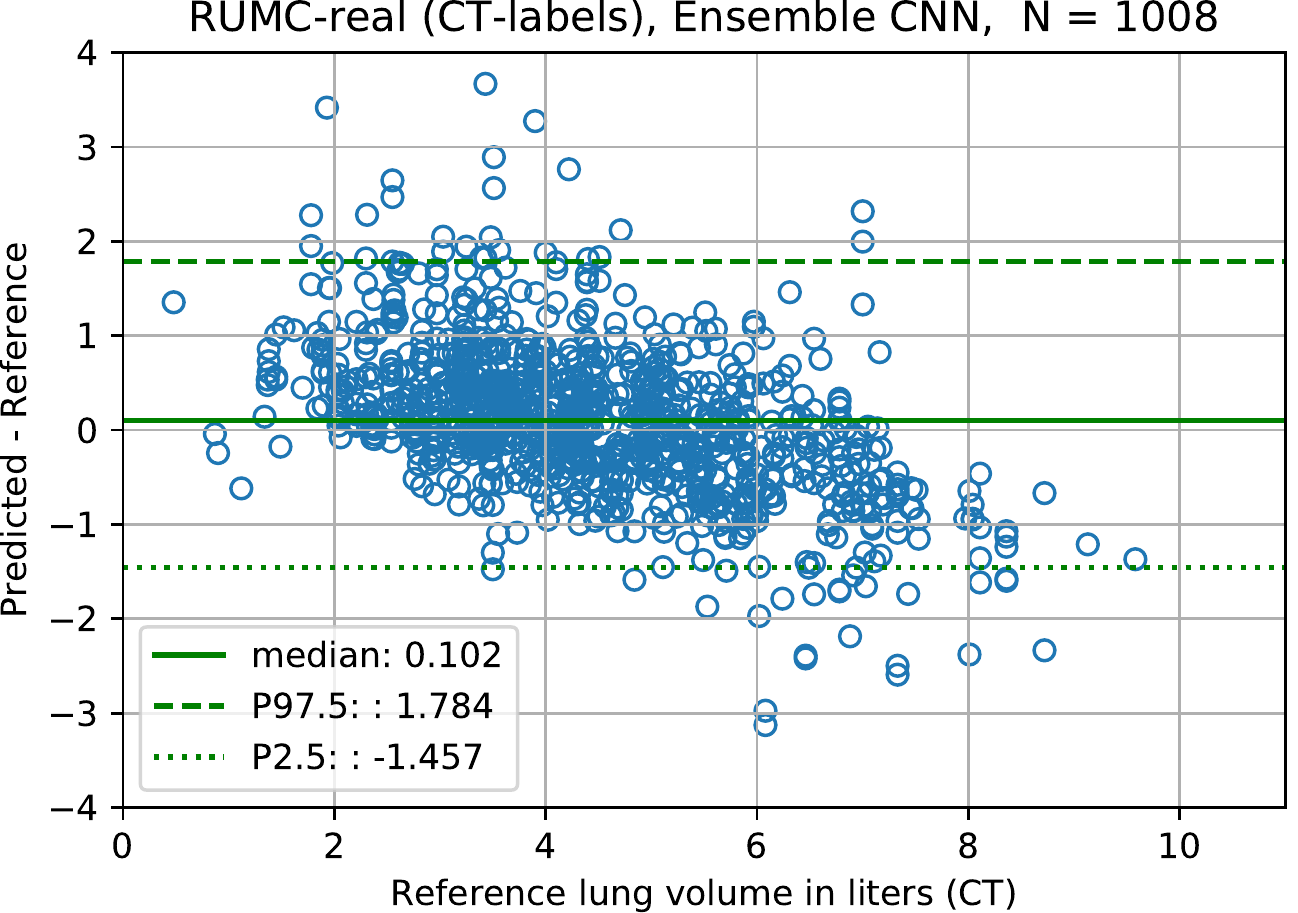}
    \caption{}

    \label{fig:histsecond}
\end{subfigure}

\begin{subfigure}{0.49\textwidth}
\centering
    \includegraphics[width=\textwidth]{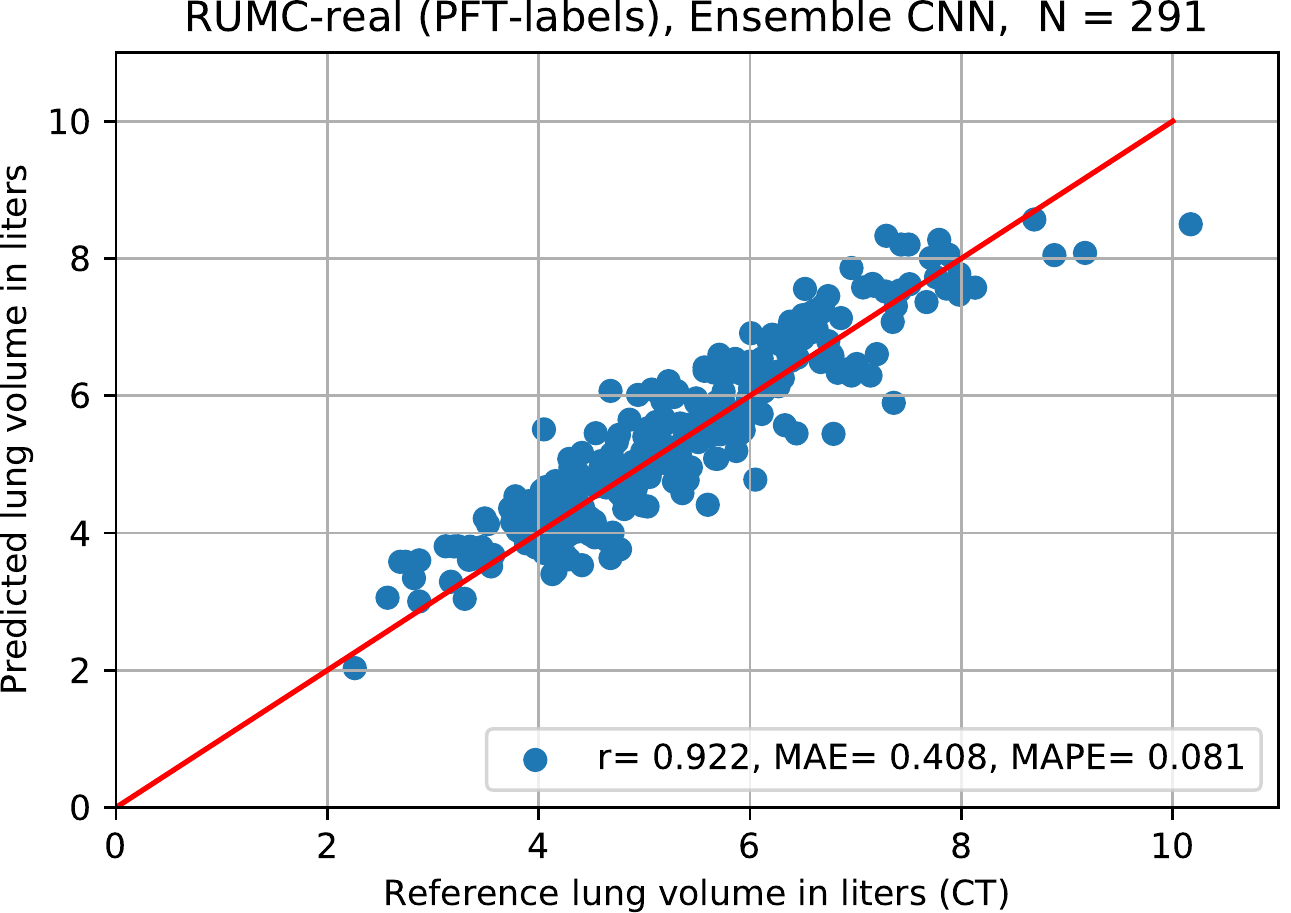}
    \caption{}

    \label{fig:histmodel}
\end{subfigure}\hfill
\begin{subfigure}{0.49\textwidth}
\centering
    \includegraphics[width=\textwidth]{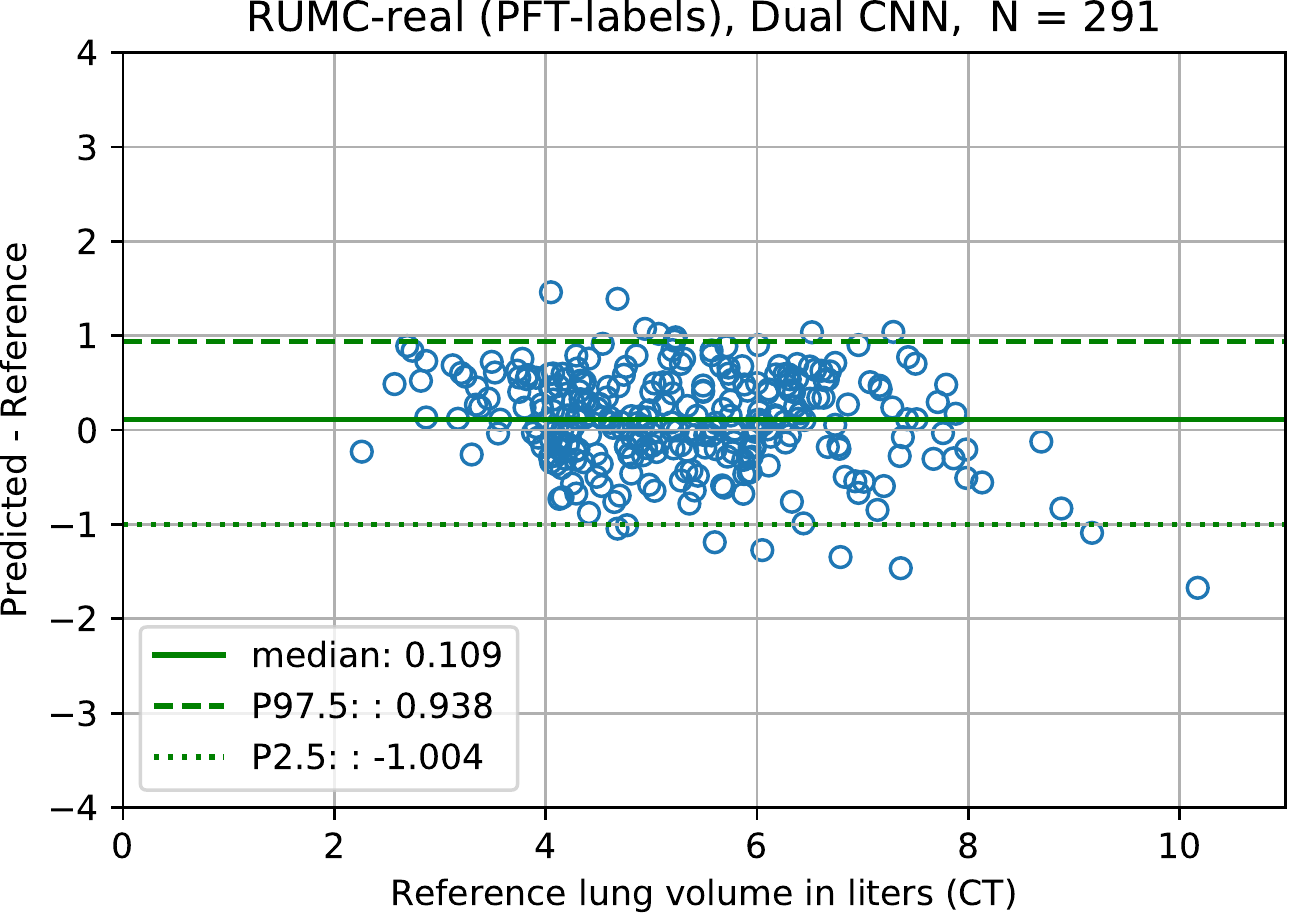}
    \caption{}

    \label{fig:histsecond}
\end{subfigure}
\caption{Results on real datasets in step-wise experiments. Left: The TLV predictions of the best model against the reference standard measurements on the held-out evaluation sets. (a) RUMC-real, (c) RUMC-real (PFT-labels). Red line is line of identity (ideal agreement). Right: Bland-Altman-like plot to analyze the differences between predicted and reference standard TLV measurements. Non-parametric method was used to estimate 95\% limits of agreement. Abbreviations: r = Pearson correlation coefficient, MAE = mean absolute error, MAPE = mean absolute percentage error, N = number of data, P2.5 = 2.5th percentile P97.5= 97.5th percentile.}
\label{fig:anaylsis}
\end{figure*}

In the COPDGene-sim dataset, where chest radiographs were simulated from CT and potential sources of label error were minimal, VGG-Net, 6-layer CNN, and Densenet121 architectures were selected. On the held-out evaluation set the model with the lowest error according to all 3 metrics was the dual CNN with 6-layer CNN architecture. This model achieved a mean absolute percentage error (MAPE) of 2.2\% and mean absolute error (MAE) of only 112ml. The scatter plot of model predictions against the reference standard from CT volumes and Bland-Altman-like plot for analyzing differences between the reference standard and predicted TLV measurements are shown in Figure \ref{fig:anaylsis} (a) and (b), respectively. As shown in Figure \ref{fig:anaylsis} (b), 95\% of differences between predicted and reference standard TLV were from -351ml to 261ml.

On the RUMC-sim dataset, which contains more abnormal images compared to COPDGene-sim, Densenet121, and ResNet architectures were selected from the validation set experiments. As in the COPDGene-sim experiments, the lateral CNN model performed better than the frontal CNN model and the best performance on the evaluation set was, once again, achieved by the dual CNN with MAPE of 2.9\% and MAE of 112ml as seen in Table 2 and plotted in Figure \ref{fig:anaylsis} (c). Limits of agreement of the differences between the predicted and reference standard TLV measurements was between -348ml to 235ml as shown in Figure \ref{fig:anaylsis} (d).
 
\begin{figure*}[h!]
\centering
\begin{subfigure}{0.49\textwidth}
\centering
    \includegraphics[width=\textwidth]{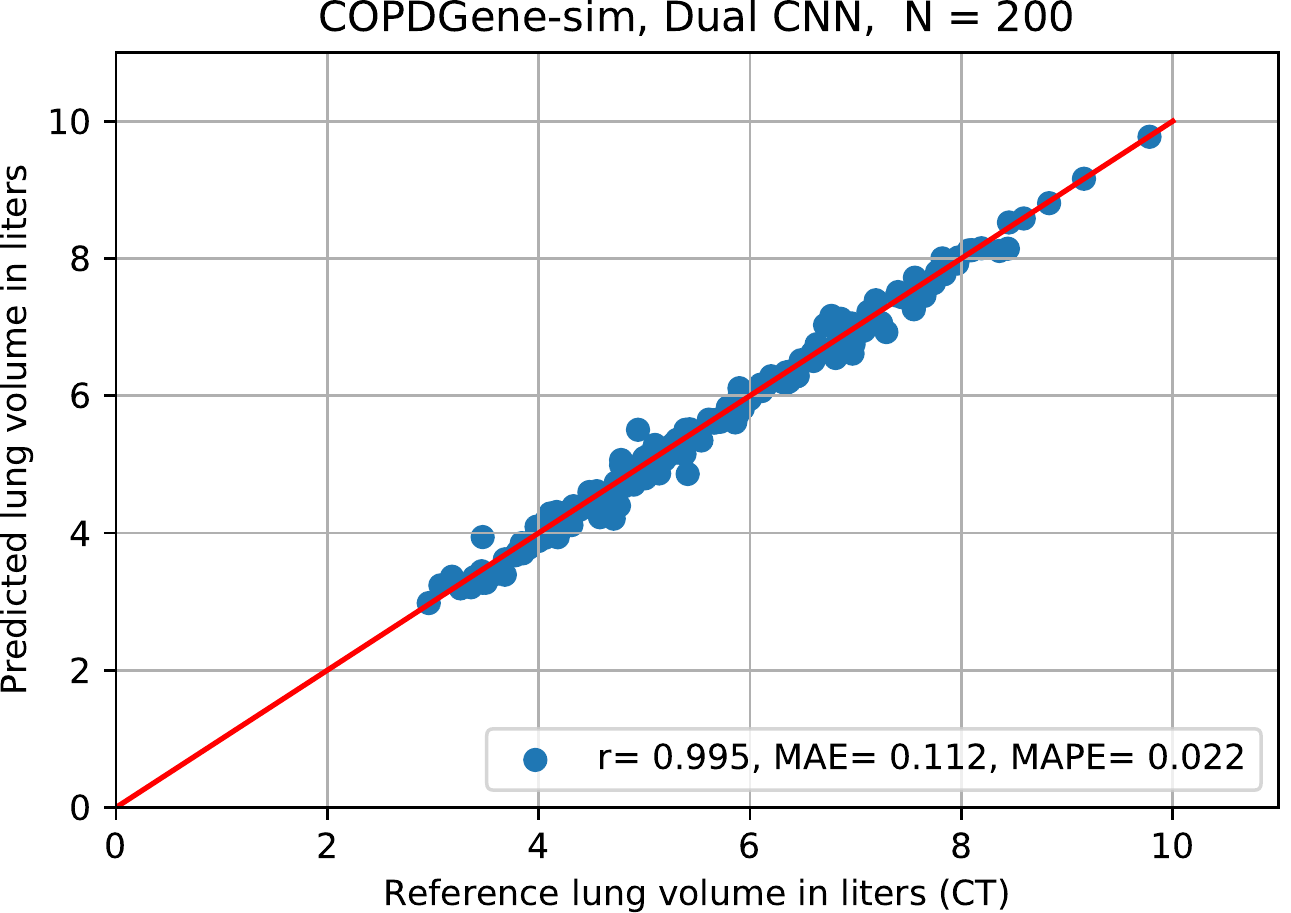}
    \caption{}

    \label{fig:scattermodel}
\end{subfigure}\hfill
\begin{subfigure}{0.49\textwidth}
\centering
    \includegraphics[width=\textwidth]{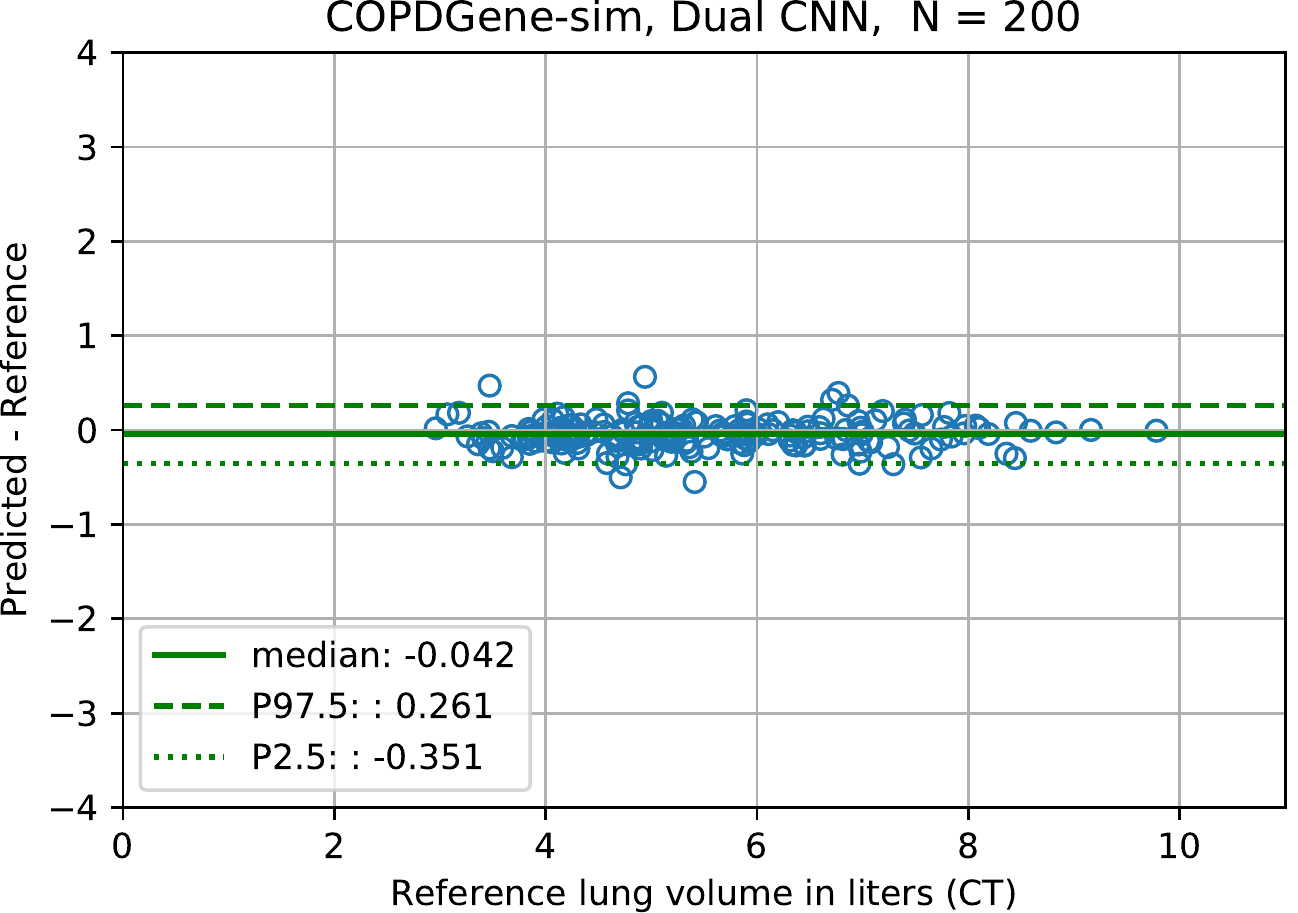}
    \caption{}

    \label{fig:scattersecond}
\end{subfigure}

\begin{subfigure}{0.49\textwidth}
\centering
    \includegraphics[width=\textwidth]{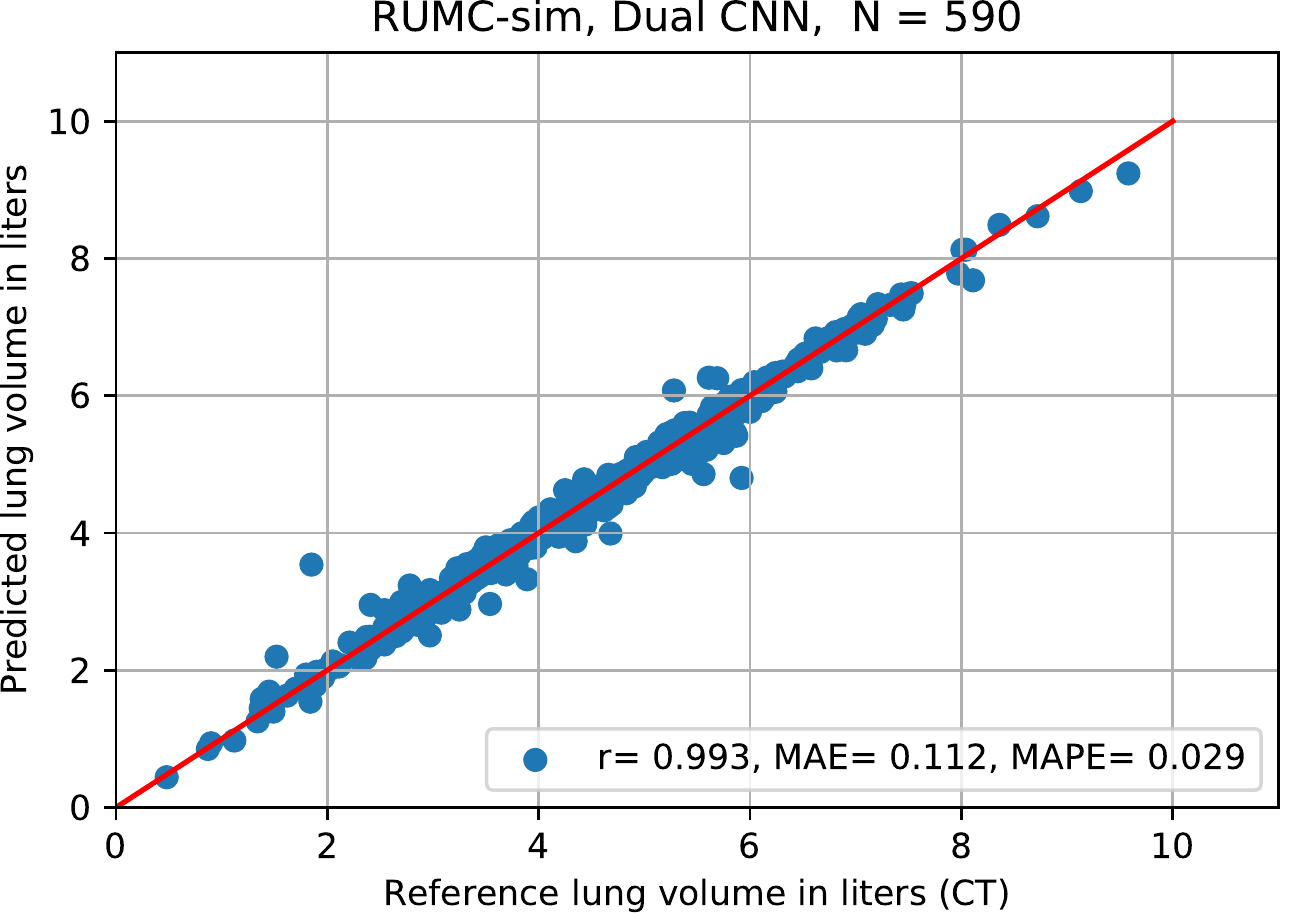}
    \caption{}

    \label{fig:scattermodel}
\end{subfigure}\hfill
\begin{subfigure}{0.49\textwidth}
\centering
    \includegraphics[width=\textwidth]{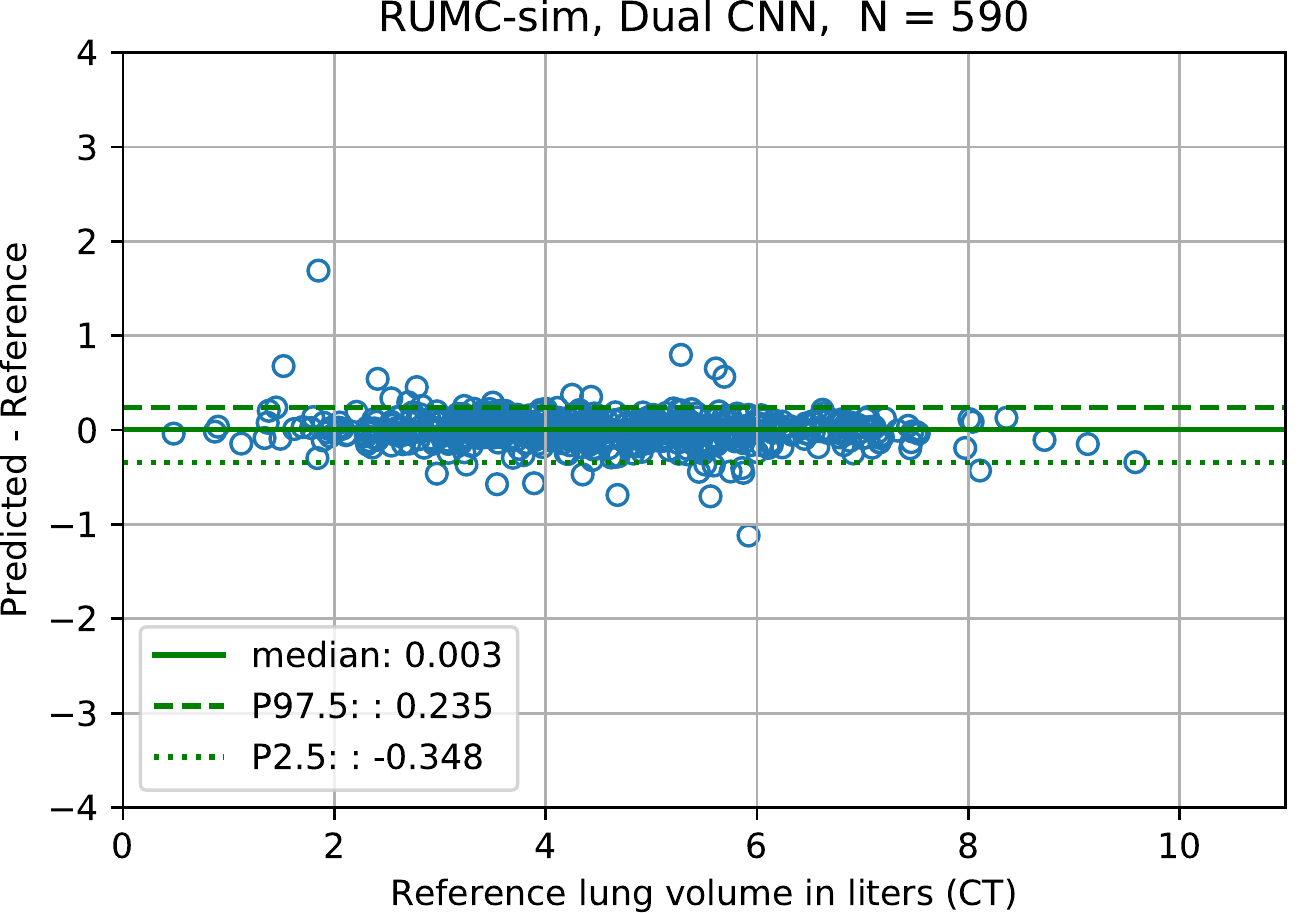}
    \caption{}

    \label{fig:scattersecond}
\end{subfigure}
    \caption{Results on simulated datasets in step-wise experiments. Left: The TLV predictions of the best model against the reference standard measurements on the held-out evaluation sets. (a) COPDGene, (c) RUMC-sim. Red line is line of identity (ideal agreement). Right: Bland-Altman-like plot to analyze the differences between predicted and reference standard TLV measurements. Non-parametric method was used to estimate 95\% limits of agreement. Abbreviations: r = Pearson correlation coefficient, MAE = mean absolute error, MAPE = mean absolute percentage error, N = number of data, P2.5 = 2.5th percentile P97.5= 97.5th percentile. }
\label{fig:anaylsis_2}
\end{figure*}

Finally, in the RUMC-real dataset, where real chest radiographs were used, dual CNN and ensemble CNN performed very similarly, and the best result obtained (with the least MAPE) with CT-derived labels was achieved by the ensemble CNN, as shown in Table 2. This model achieved 15.7\% MAPE, and MAE of 597ml. The model predictions and references for the evaluation set of 1008 CXRs are plotted in Figure \ref{fig:anaylsis_2} (a); and the differences between predicted TLV and reference standard is analyzed in Figure \ref{fig:anaylsis_2} (b). As shown in Figure \ref{fig:anaylsis_2} (b), the model tended to underestimate TLV where reference standard was higher than 6 liters, and overestimate TLV where reference standard was lower than 4 liters.

For the final experiment using PFT-derived labels, the best models trained on the RUMC-real (CT-labels) data for frontal, lateral, dual CNN were used as pretrained models and further fine-tuned on 637 CXR images with PFT-derived labels. The results achieved on 291 CXR images with PFT-derived labels are shown in in Table 2 (RUMC-real (PFT-labels)). The best model on the held-out evaluation set was the dual CNN with ResNet34 architecture and achieved MAE of 408ml and MAPE of 8.1\%. The model predictions and PFT-derived reference standard were highly correlated with Pearson correlation coefficient of 0.92 as illustrated in Figure \ref{fig:anaylsis_2} (c); 95\% of differences between predicted and reference standard TLV measurements were from -1 liters to 938 ml (Figure \ref{fig:anaylsis_2} (d)). 

\begin{figure}[ht]
    \centering
    \begin{subfigure}{0.49\textwidth}
\centering
    \includegraphics[width=\textwidth]{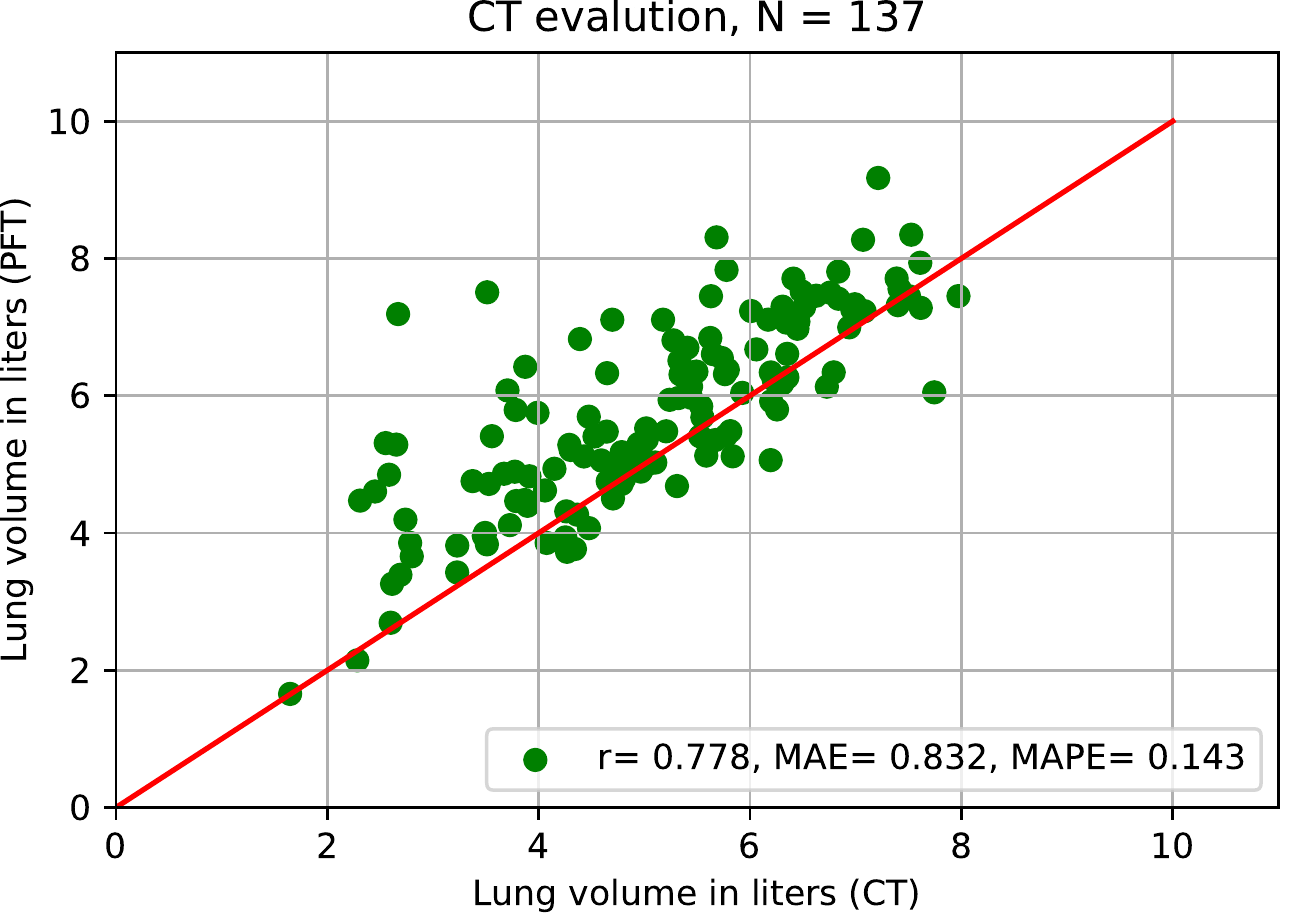}
  
    \caption{}

    \label{fig:scatter_ct_eval}
\end{subfigure}\hfill
\begin{subfigure}{0.49\textwidth}
\centering
    \includegraphics[width=\textwidth]{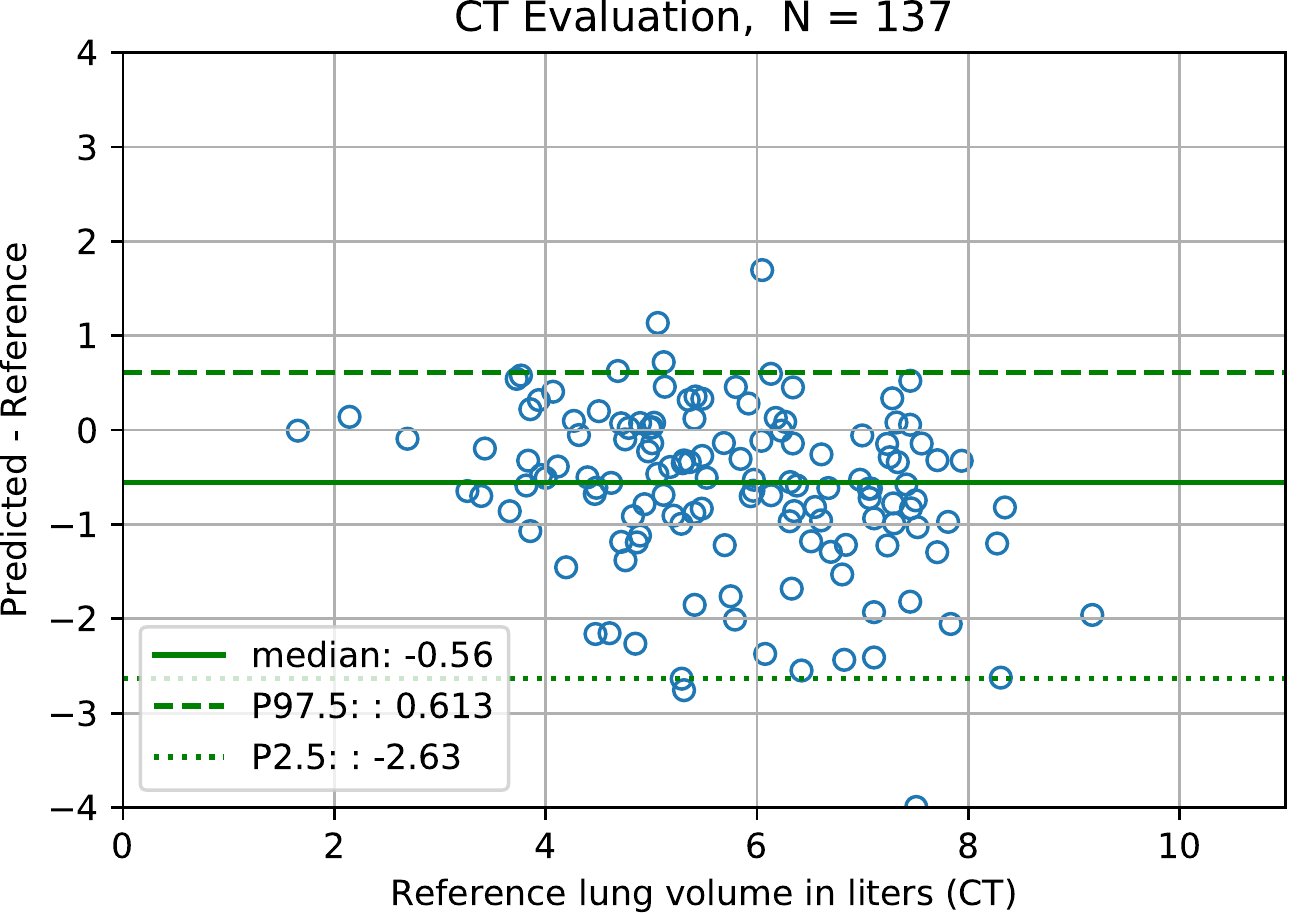}
    \caption{}

    \label{fig:histsecond}
\end{subfigure}

    \caption{CT-derived TLV against PFT-derived TLV on CT evaluation dataset. Left: Comparison of CT-derived total lung volumes with Pulmonary Function Test total lung volumes on the CT evaluation set. Right: Bland-Altman-like plot to analyze differences between CT-derived and PFT-derived total lung volume. Abbreviations: r = Pearson correlation coefficient, MAE = mean absolute error, MAPE = mean absolute percentage error, N = number of data, PFT = pulmonary function test, P97.5 = 97.5th percentile, P2.5 = 2.5th percentile. }
    \label{fig:ct_evaluation}
\end{figure}

Figure \ref{fig:ct_evaluation} (a) and (b) shows the results of the comparison between CT-derived TLV and PFT-derived TLV on the CT evaluation set of 137 subjects. These two measurements were well correlated with Pearson's r of 0.78, however, considerable variations were observed between the two measurements for some patients. TLV was consistently underestimated by CT-based measurements where median differences (bias) between CT-derived and PFT-derived was -560ml as shown in Figure \ref{fig:ct_evaluation} (b).

\begin{table}[ht]
\centering
\begin{tabular}{ |l|l|l|l|l|l|}
\hline
 Evaluation Datasets (\#images)& Model & Architecture & MAPE(\%) & MAE(ml) & Pearson's r \\
\hline
\multirow{4}{*}{COPDGene-sim (200)} & Frontal CNN & DenseNet121 & 4.3 & 226 & 0.978 \\
 & Lateral CNN & VGG-Net & 3.6 & 198  & 0.983 \\
 & \textbf{Dual CNN} & \textbf{6-layer CNN} & \textbf{2.2} & \textbf{112} & \textbf{0.995} \\
 & Ensemble CNN & DenseNet121\&VGG-net & 2.6 & 139 & \textbf{0.992}\\ \hline
 \multirow{4}{*}{RUMC-sim (590)} & Frontal CNN & DenseNet121 & 5.5 & 220 & 0.978 \\
 & Lateral CNN & DenseNet121 & 5.0 & 200 & 0.984\\
 & \textbf{Dual CNN} & \textbf{ResNet34} & \textbf{2.9} & \textbf{112} & \textbf{0.993}\\
 & Ensemble CNN & DenseNet121\& DenseNet121 &3.8 & 154 & 0.989\\ \hline
 \multirow{4}{*}{RUMC-real (CT-labels) (1008)} & Frontal CNN &VGG-Net & 16.9 & 650  & 0.826 \\
 & Lateral CNN & DenseNet121 &16.8 & 639  & 0.831\\
 & Dual CNN & ResNet34 &16.1 & \textbf{592} & \textbf{0.855}\\
 & \textbf{Ensemble CNN} & \textbf{VGG-Net \& DenseNet121} & \textbf{15.7} & 597 & 0.851\\ \hline
\multirow{4}{*}{RUMC-real (PFT-labels) (291)} & Frontal CNN &VGG-Net & 10.3 & 509  & 0.870 \\
 & Lateral CNN & DenseNet121 &9.2 & 472  & 0.875\\
 & \textbf{Dual CNN} & \textbf{ResNet34} & \textbf{8.1} & \textbf{408} & \textbf{0.922}\\
 & Ensemble CNN & VGG-Net \& DenseNet121 & 8.5 & 420 & 0.907\\ \hline

\end{tabular}
\label{tab:mae}
\caption{Results of the selected models on the held-out evaluation sets. Mean absolute error is calculated against the reference standard for TLV measurements. MAE = mean absolute error (in milliliters), MAPE = mean absolute percentage error, Pearson's r = Pearson correlation coefficient. Bold font indicates best performance per dataset and metric.}
\label{tab:mae}
\end{table}

\section*{Discussion}
This study demonstrated that state-of-the-art deep learning solutions can measure TLV from PA and lateral CXRs, using primarily CT-derived labels and a small number of PFT-derived measures. To demonstrate the sources of error, the experiments were conducted in a step-wise fashion with increasing levels of complexity.  Using simulated CXRs eliminated potential error related to the patient position or inspiration level between the CT and CXR image acquisition. Results on both simulated datasets show extremely low error (MAPE of 2.2\% and 2.9\%) and high correlation with the reference labels (r=0.99 and r=0.99).  The slightly better performance on the COPDGene-sim dataset may be attributed to the fact that this dataset contains a limited range of pathologies and that the CT segmentations were manually corrected, meaning that even very small inaccuracies were eliminated.  

In the dataset of clinical CXR with CT-derived volumes (RUMC-real dataset) we see a substantial increase in the prediction error with MAPE of 15.7\%, which we attribute largely to the difference in patient position and inspiration effort between the CT and the CXR image acquisition. It is likely that the degree of inspiration in the CXR and CT images is different, particularly given that there is known to be a high intra-individual deviation in TLV between routine CT scans (\cite{Haas14}). The indication from this experiment is that CT-derived labels are useful, but not optimal, to learn the TLV from CXR. As an additional check we investigated the relationship between CT-derived and PFT-derived volumes in 137 cases where both were available. This provides results in line with previous studies on CT-derived lung volumes~\cite{Haas14,Dagh19}: although CT-derived lung volume and TLV are well correlated (r=0.78), there are considerable differences in some patients. 

To overcome the issues with the CT-derived labels on the RUMC-real dataset we further fine-tuned the best networks from that experiment with PFT-derived labels. Evaluation on an independent dataset of 291 subjects that were not used for training showed that the error of the estimated TLV from CXR relative to the measured TLV from PFT is reduced considerably, achieving MAPE of 8.1\% and Pearson's correlation coefficient of 0.92. This algorithm is publicly available at “https://grand-challenge.org/algorithms/cxr-total-lung-volume-measurement/”.

In all experiments the model was optimized to use the best performing architecture and input. In the experiments using simulated CXR images, it is notable that the networks using lateral images as input perform better than the networks using frontal images. This may indicate that the lateral projection image contains more information related to CT-derived TLV. However we note also that in all experiments the combination of frontal and lateral images produced the optimal results, either by use of a dual-CNN or through an ensemble. 

Previous literature has investigated predictive equations for measurement of TLV from chest radiographs using manual measurements. One study \cite{Park16} investigated performance with simulated chest radiographs to predict CT-derived TLV. Their method, which required manual measurements, had an inferior performance (MAPE of 5.7\%) on their dataset compared to our results obtained in the COPDGene-sim and RUMC-sim datasets (MAPE of 2.2\% and 2.9\%). For studies which investigated predictive equations to estimate PFT-derived TLV from real CXR \cite{Pier79, Harr71, Barn60}, the coefficient of correlation between predictions and reference standard (body plethysmography or helium dilution technique) generally ranged from 0.80 to 0.93 (compared to our method with 0.92). Sample sizes in these papers ranged from 21 to 100 patients. However, it should be noted that many of these studies used spirometric control to regulate the level of inspiration during CXR acquisition. In fact, one study \cite{Wade51} has shown that without spirometric control the correlation of predicted TLV and PFT-derived reference standard was only 0.47, compared to 0.82 with spirometric control. In this work, however, we experiment with routinely taken chest radiographs (with no spirometric control), and produce TLV predictions which are highly correlated (r=0.92) to PFT-derived results. Our work is the first to demonstrate automated measurement of TLV from chest radiographs and achieves a comparable or lower error range with a remarkably larger sample size compared to previous literature. 

There are several limitations in this study. First, the datasets were constructed from routinely taken studies with the assumption that TLV would not change in 15 days, which might not hold true for extreme cases. This selection criterium also yielded an under-representation of healthy subjects but reflects a clinical population in which TLV measurements are of clinical interest. The PFT-derived reference standard measurements were obtained using the helium dilution technique which might underestimate TLV in patients with severe airway obstruction. Furthermore, inspiration levels were not controlled in a similar fashion to PFT in these routine chest radiographs, which could have introduced a source of error in our predictions, but this represents regular clinical practice. One possible solution to address this issue would be to develop an automated algorithm to assess the inspiration level on CXR, for example by rib counting \cite{Made18}. Moreover, our held-out evaluation set was constructed with patients assessed for lobectomy since their PFT results were readily available; future research should address the evaluation of the algorithm on a population with other clinically relevant pathologies, including fibrosis.

In conclusion, we demonstrated that TLV can be automatically estimated from CXR using a deep-learning approach, with an accuracy that is superior or comparable to the previous literature using semi-automated methods. Further, we showed that the deep learning system can be trained primarily with CT-derived labels from automatically segmented chest CT images, and fine-tuned on gold-standard PFT-derived labels. This automated system could be routinely applied to clinical chest radiographs and serve as a tool for identifying temporal change in total lung volume in patients with restrictive and obstructive lung diseases.

\bibliography{sample}

\section*{Acknowledgements}
We thank Weiyi Xie for his help in dataset collection, and Jesus Lago for useful discussion on statistical analysis. This work was supported by the Dutch Technology Foundation STW, which formed the NWO Domain Applied and Engineering Sciences and partly funded by the Ministry of Economic Affairs (Perspectief programme P15-26 ‘DLMedIA: Deep Learning for Medical Image Analysis’). 
\end{document}